\documentclass[12pt]{article}
\usepackage[pdftex]{graphicx}
\usepackage{caption}
\usepackage{wrapfig}
\usepackage{latexsym}
\usepackage[pdftex]{color}
\usepackage{pifont}
\usepackage{epstopdf}
\usepackage{tikz}
\usepackage{mathtools}
\usetikzlibrary{decorations.pathmorphing,patterns}

\newcommand{\bmp}[2][t]{\begin{minipage}[#1]{#2}}
\newcommand{\emp}{\end{minipage}}

\usepackage{amsmath}
\usepackage{amssymb}
\usepackage{textcomp}
\usepackage{tcolorbox}
\usepackage[title]{appendix}

\usepackage{sidecap}

\usepackage[square,numbers,sort&compress]{natbib} 

\newcommand{\bblubox}[1]{\begin{tcolorbox}[colframe=blue!75!white,title=#1]}
\newcommand{\eblubox}{\end{tcolorbox}}

\usepackage{lineno}

\begin{document}
\title{Effectiveness of Outer Hair Cells as Cochlear \\ Amplifier:
In Simple Model Systems}
\date{}
\author{
Kuni H Iwasa\\
\normalsize{NIDCD, National Institutes of Health}\\
\normalsize{Bethesda, MD 20892, USA}
}

\maketitle

\abstract{Cochlear outer hair cells (OHCs) have two mechanosensitive elements:  the hair bundle with mechanotrasducer channels and the piezoelectric lateral wall of the cell body. The present report examines how these elements interact with each other by incorporating OHCs into the simplest local cochlear models.
In the frequency range, typically above 1 kHz, where capacitive conductance is greater than the ionic conductance, hair bundle (HB) conductance drives the piezoelectric cell body and amplified oscillation by countering viscous drag, while the cell body increases its stiffness owing to strain-induced polarization, elevating the resonance frequency.  Since HB sensitivity is essential for amplification, the resonance is not pure piezoelectric, but semi-piezoelectric. In the lower frequency range, typically lower than 100 Hz, strain induced polarization  contributes to drag and the HB sensitivity increases cell body stiffness. }

\textbf{keywords:} \emph{hearing, sensitivity, piezoelectricity, mechanics}

\section*{Statement of Significance}
Outer hair cells are essential for the sensitivity, frequency selectivity, and the wide dynamic range of the mammalian ear. For this reason, these cells have been extensively studied for over 30 years using isolated cell preparations. These studies revealed not only the properties of the ion channels and the mechanosensitive hair bundle, but novel piezoelectricity in the cylindrical cell body. This paper addresses the question as to how these properties of outer hair cells are integrated in the ear to function as the cochlear amplifier and enhances the performance of the mammalian ear. More specifically, this study examines the performance of outer hair cells by incorporating them into the simplest mechanical models for the mammalian cochlea. 

\section*{Introduction}
Outer hair cells (OHCs) are essential for the sensitivity of the mammalian ear for providing mechanical feedback to amplify the vibration in the cochlea\cite{ld1984,Dallos2008}. These cells have two mechanosensitive mechanisms:  the piezoelectric lateral wall\cite{bbbr1985,a1987} and the hair bundle\cite{hh1987}. The present report studies how the interplay of these two mechanisms determines their response to external mechanical stimulation.

The mechanosensitivity of the mammalian hair bundle \cite{hh1988,zshlmd2000,ric-fet2003,BeurgRicci2009} and the motile mechanism based on piezoelectricity in the cell body \cite{a1990,i1993,ia1997,Santos-Sacchi1998a,ai1999,zshlmd2000,SantosNava2023} have been carefully studied separately, both experimentally and theoretically. Studying the interplay of these two mechanisms is more complex because it depends on the connectivity of the OHC to its mechanical environment, particularly on the way the motile response of the OHC is fed back on the activation of its hair bundle. 

Here an OHC is incorporated into the two simplest mechanical systems to examine its motile activity study to clarify the interplay of the two mechanosensitive elements. More specifically, the present study focuses on the displacement and stiffness changes of the cell  during forced periodic oscillation.

\section*{Basic equations}
In the following, we consider each of the two simplest mechanical systems, where an OHC is incorporated. To describe these systems, let us start from piezoelectricity of the lateral wall, and then the equation of motion, followed by membrane currents in the OHC. 

\subsection*{Piezoelectric lateral wall} 
\begin{SCfigure}[2] [b]
\includegraphics[width=0.32\linewidth]{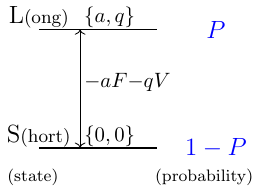} 
\caption{\small{Two states, long and short, of the motile element. The probability of state L is $P$ and that of state S is $1-P$. Energy level of state L is higher than that of S by $-aF-qV$ when the membrane potential is $V$ and the force applied to the motile element is $F$. Notice that $a>0$ and $q<0$  because depolarization shrinks the cell.
}}
\label{fig:2states}
\end{SCfigure}

Assume that the motile elements in the lateral membrane of the OHC have two states, the long state $L$ and the short state $S$, and $P$ is the fraction of the motile element in state $L$ (Fig.\ \ref{fig:2states}). The natural length of the cell is $X_0+aNP$, where $a$ is the contribution of a single motile element to the cell length associated with conformational change from $S$ to $L$. This conformational change accompanies movement of charge $q$ across the membrane. The number of the motile elements in the cell is $N$. In the equilibrium condition, $P$ is determined by the free energy $G$, which is the sum of an electrical term $qV$ and a mechanical term $aF$, where $V$ is the membrane potential (or voltage) and $F$ the force applied to the cell due to electromechanical coupling in the absence of external load \cite{i1994,i1990}.

Now put mechanical load, including an external elastic element $K_e$, a viscous element  with drag coefficient   $\eta$, and a mass $m$, on an OHC with intrinsic stiffness $k_o$  (Fig.\ \ref{fig:mech_schem}). Assume that this system is initially in equilibrium, and then change the membrane potential by $\Delta V$. The motile element undergoes conformational change {that results in}    $\Delta P$. The electrical component of the free energy change is $q\Delta V$. The resulting mechanical displacement $\Delta X_p=aN\Delta P$ generates force $\Delta F_i$ due to the external elastic load. This force, in turn, produces an elastic displacement $\Delta X_e$ in the OHC, resulting in the net displacement $\Delta X=aN\Delta P-\Delta X_e$ on the OHC as well as the external elastic element  $K_e$ . Thus, $\Delta F_i=  K_e   \Delta X=k_o \Delta X_e$ is the internally produced force. These relationships leads to
\begin{subequations}
\begin{align}\label{eq:delX}
 \Delta X&=aN\Delta P\;k_o/(k_o+K_e), \\
 \Delta F_i&=aN\Delta P\; K_ek_o/ (k_o+K_e).
 \label{eq:DeltaFi}
\end{align}
\end{subequations}
The above expression for $\Delta F_i$ is valid for the two cases of connectivity in Fig.\ \ref{fig:mech_schem}A and B.

The motile machinery also responds to an externally applied force $F_\mathrm{ext}$. The force $F_e$ applied to the motile machinery depends on the connectivity (see Fig.\ \ref{fig:mech_schem}) and is given by
\begin{align} \label{eq:DeltaFe}
\Delta F_e=\left\{ \begin{array}{cl} 
F_\mathrm{ext}& \mbox{with series connectivity,} \\
k_o/(k_o+K_e)F_\mathrm{ext}& \mbox{with parallel connectivity.}
\end{array}\right. 
\end{align}
The total free energy $\Delta G$ is expressed by $ q\Delta V-a(\Delta F_i+\Delta F_e)$ with $\Delta F_i$ given by Eq.\ \ref{eq:DeltaFi} and $\Delta F_e$ given by Eq. \ref{eq:DeltaFe}.

We proceed by assuming that the free energy of the motile element at any given moment is determined by the given values of voltage $V$ and the mechanical strain $X$ of that moment, exactly the same as in the static case as described above. Then the variable $P$ of the motile element changes toward its equilibrium value $P_\infty$, which is given by the Boltzmann function
\begin{subequations}
\begin{align}
\label{eq:Pinf}
P_\infty&=\exp[-\beta\Delta G]/(1+\exp[-\beta\Delta G]) \quad \mathrm{with} \\
 \Delta G&=G_0-q\Delta V-a(\Delta F_i+\Delta F_e), 
 \label{eq:DeltaG}
\end{align}
\end{subequations}
where the free energy difference $\Delta G$ is from state $S$, the short conformation so that $P$ is proportional to $X$ and $\beta=1/(k_BT)$ with Boltzmann's constant $k_B$ and the temperature $T$. The constant term $G_0$ determines the ratio of the two states at $\Delta V=0$ and $\Delta F=0$.

The signs of the parameters are $a>0$ and $q<0$ because an increase in axial force $F$ leads to elongation and a positive shift in $V$ (depolarization) results in a reduction of the cell length (see below for details).

If the motor state $P$ satisfies $P=P_\infty$, the system is in equilibrium and does not undergo movement. The difference $P-P_\infty$ drives the system.
Here, we assume that the deviation of our system from equilibrium is small ($\beta\Delta G\ll 1$). Then, we can expand the Boltzmann function to the first order term
\begin{align}
P_\infty\approx P_\infty^0-\beta \gamma\Delta G/4
\label{eq:Pexp}
\end{align}
with $\gamma=4P_\infty^0(1-P_\infty^0)$.  The factor four is introduced so that the maximal value of $\gamma$ is unity.  

\subsection*{Equation of motion}  
\begin{SCfigure}[0.9] 
\includegraphics[width=0.5\linewidth]{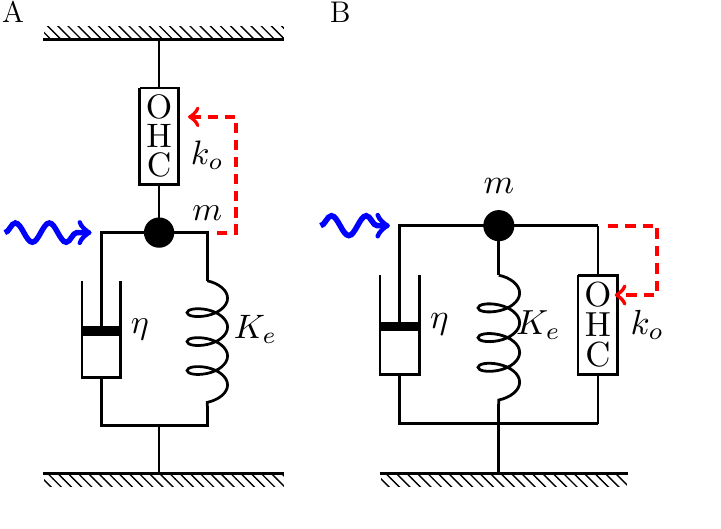} 
\caption{An OHC is incorporated into two cases of the two simplest systems with elastic load  $K_e$   and viscous load $\eta$. A: series connectivity, B: parallel connectivity. The OHC responds to the movement of mass (dashed red arrow).  These two cases lead to different values of $\alpha_c$ in Eq.\ \ref{eq:alfc_def}.}
\label{fig:mech_schem}
\end{SCfigure}

Now an OHC is incorporated into two cases of the simplest mechanical system (Fig.\ \ref{fig:mech_schem}).
The equation of motion of the system can be formally expressed by
\begin{align}
 m \; d^2X/dt^2+\eta \; dX/dt=k_o(X_\infty-X)+F_\mathrm{ext},
 \label{eq:x_eqn}
\end{align}
where $X_\infty=aNP_\infty\; k_o/(k_o+K_e)$, which shares the same factor that relates $X$ and $P$, is the displacement that corresponds to the equilibrium for the present set of values for force $F_i+F_e$ and voltage $V$. The difference between the present displacement $X$ and $X_\infty$ drives the system.

Here $m$ is the mass, $\eta$ drag coefficient, and $F$ external force. The inertia term can be justified if the system is not far from equilibrium \cite{Iwasa2021}.
This equation can be expresses using variable $P$ by dividing both sides with $aN k_o/(k_o+K_e)$
\begin{subequations}
\begin{align}  \label{eq:p_eqn1}
m \; d^2P/dt^2+\eta dP/dt&=(k_o+K_e)(P_\infty-P)+F \quad \mathrm{with} \\
\label{eq:Fext}
F&=F_\mathrm{ext}(k_o+K_e)/(aNF k_o).
\end{align}
\end{subequations}

Now consider the response to a sinusoidal voltage waveform of small amplitude $v$ and angular frequency $\omega$. Let $p$ be the corresponding small amplitude of $P$.
\begin{align*}
F=F_0+f\exp[i\omega t], \quad
V=V_0+v\exp[i\omega t], \quad
P=P_0+p\exp[i\omega t].
\end{align*}
The amplitude $x$ of displacement is related to $p$ with
\begin{align}
  x=aNp\;k_o/(k_o+K_e).
  \label{eq:x_p}
\end{align}

By using the linearized $P_\infty$ (i.e.\ Eq.\ \ref{eq:Pexp}), the equation of motion (\ref{eq:p_eqn1}) is transformed into
\begin{align}
[-\overline\omega^2+i\overline\omega/\overline\omega_\eta+1] p=(1/4)\beta\gamma [q v+a(f_i+f_e)]+\overline{f},
\label{eq:p_omega}
\end{align}
after dividing both sides with $k_o+K_e$ and using a new reduced frequency $\overline\omega(=\omega/\omega_r)$, normalized to the mechanical resonance frequency $\omega_r=\sqrt{(k_o+K_e)/m}$. Here, $\overline\omega_\eta$ is normalized viscoelastic roll-off frequency, $\gamma=4 P_0(1-P_0)$. With the definition $\overline{f}=f/(k_o+K_e)$ and Eq.\ \ref{eq:Fext}, $\overline f$ is dimensionless, being the amplitude of external oscillatory force divided by $aNk_o$. The quantity $f_i$ and $f_e$ are the amplitudes of internally induced force and externally induced force (Eq.\ \ref{eq:DeltaFe}), respectively. 

The parameter $\gamma$ originates from the expansion of the exponential term of $P_\infty$. It takes the maximal value of unity at $P_0=0.5$, where the motile mechanism is the most sensitive. For this reason, $\gamma$ could be called ``operating point parameter'' or ``activity of the electromotile (em) mechanism.'' 

The right-hand-side of Eq.\ \ref{eq:p_omega} can be expressed as
\begin{align}
\frac{\beta\gamma}{4} qv+\gamma u_a \overline K_e p+(\alpha_c\gamma u_a+1)\overline{f},
\label{eq:eom_p}
\end{align}
with notations defined by
\begin{subequations}
\label{eq:KeUaAlf}
\begin{align} \label{eq:ua_def}
u_a&=\beta a^2Nk_o, \\  \label{eq:Ke_def}
\overline K_e&=\frac{K_e}{k_o+K_e},\\  \label{eq:alfc_def}
\alpha_c&=\left\{ \begin{array}{cl} 
1& \mbox{for series connection} \\
 1-\overline K_e & \mbox{for parallel connection}
\end{array}\right. 
\end{align}
\end{subequations}
The distinction in connectivity is insignificant if a condition $u_a\ll 1$ is satisfied. That will be examined later by numerical examination.

\subsection*{Current equation} 
The effect of hair bundle resistance $R_a$ on the membrane potential $V$ can be expressed
\begin{align}
 (e_{ep}-V)/R_a=(V-e_K)/R_m+C_0\; dV/dt+Nq\; dP/dt,
 \label{eq:Ra}
\end{align}
where $e_{ep}$ is the endocochlear potential, $e_K$ is the resting potential of OHC, which is primarily determined by K$^+$ conductance, and $R_a$ hair bundle  resistance  . The last term in the right-hand-side of the equation is due to the change of the motile mechanism. 

For periodic stimulation with angular frequency $\omega$, introducing the time independent component $R_0$ and the relative amplitude $\hat r$ of the hair bundle resistance $R_a$, we obtain
\begin{align}
 -i_0 \hat r=(\sigma+i\omega C_0)v+i\omega Nqp.
 \label{eq:c_eqn}
\end{align}
Here $i_0=(e_{ep}-e_K)/(R_0+R_m)$ is  the steady state current and $\sigma=1/R_0+1/R_m$ the steady state conductance. Eq.\ \ref{eq:c_eqn} shows two asymptotic behaviors: low frequency $\omega\ll\omega_m$ and high frequency $\omega\gg\omega_m$, introducing a characteristic middle frequency $\omega_m(=\sigma/C_0)$.

\section*{Stiffness and force production} 
Now the system can be described by a single equation.
By replacing the voltage amplitude $v$ in Eq.\ \ref{eq:p_omega} by using Eq.\ \ref{eq:c_eqn}, we obtain
\begin{align}
\left[-\overline\omega^2+i\overline\omega/\overline\omega_\eta+1-\gamma u_a \overline K_e \right] p
=-\frac{\beta\gamma q}{4} \frac{i\omega Nqp+i_o\hat{r}}{\sigma+i\omega C_0}+(1+\alpha_c\gamma u_a)\overline{f},
\label{eq:complex_eom1}
\end{align}
where $u_a$ and $\overline K_e$ are defined earlier in Eq.\ \ref{eq:KeUaAlf}.

Now to close the feedback loop, we assume that the change in hair bundle {resistance}   $\hat r$ is elicited by the bending of the hair bundle, which is proportional to displacement $x$, and is described by the relationship $\hat r=g_xx$,
with the sensitivity $g_x$ to displacement $x$. Since $p$ is proportional to displacement $x$, we can put $\hat r=gp$ with
\begin{align}
g=aN(1-\overline K_e)g_x.
\label{eq:g}
\end{align}
While Eq.\ \ref{eq:complex_eom1} is not very complicated, the dependence of the function on the parameters can be made much more transparent by introducing approximate expressions for low and high frequency regions.

\subsection*{High frequency approximation}

If the frequency $\omega$, is high enough to satisfy $\sigma/\omega C_0\ll 1$ , $1/(\sigma+i\omega C_0)$ can be replaced by $(1/i\omega C_0)(1-\sigma/i\omega C_0)$.
Then Eq.\ \ref{eq:complex_eom1}  can be rewritten as
\begin{align}
[-\overline\omega^2+i(\overline\omega/\overline\omega_\eta- \gamma A_h/ \overline\omega)+1+ \gamma B_h] p=(1+\alpha_c\gamma u_a)\overline{f},
\label{eq:HFreq}
\end{align}
using an anti-damping factor $A_h$ and a stiffness factor $B_h$, which are respectively defined by
\begin{subequations}
\label{eq:HFreqAB}
\begin{align}
A_h&= \frac{\beta q}{4 \omega_r C_0}\left( i_0 g-\frac{\sigma q}{C_0}\right),\\
B_h&=-u_a \overline K_e+\frac{\beta Nq}{4 C_0}\left(q+\frac{\sigma a g i_0}{\omega_r^2\overline\omega^2C_0}\right).
\end{align}
\end{subequations}
Here, the second term in the curved brackets of $B_h$ shows dependence on the variable $\overline\omega$. However, this term can be ignored because it includes the small factor $\sigma/\omega C_0$ (see the numerical section for confirmation). 

The second term of $B_h$ is positive and increases with $\gamma$, piezoelectric activity. The source of this term is the last term in Eq.\ \ref{eq:c_eqn}, which is polarization induced by strain.  This effect can be called ``strain-induced polarization stiffness,'' a result of piezoelectric activity in the OHC. The factor $B_h$ increases the resonance frequency of the system by increasing the stiffness of OHC if the second term is larger than the first term.

The coefficient $A_h$ works as an anti-damping term if the condition $A_h>0$ holds. A requirement for this is $g<0$  because $q<0$. This condition appears intuitive. A decrease $p$, which is shortening of the OHC, is likely associated with a basilar membrane movement toward the tectorial membrane, which results in HB bending and an increase transducer current. 

For the OHC to be effective as an amplifier, the factor $A_h$ requires to satisfy an additional condition. Eq.\ \ref{eq:HFreq} leads to
\begin{align}
|p|^2=\frac{(1+\alpha_c\gamma u_a)^2\overline{f}^2}{(1+\gamma B_h-\overline\omega^2)^2+(\overline\omega/\overline\omega_\eta-\gamma A_h/ \overline\omega)^2},
\end{align}
which implies that, at $\gamma=1$, $A_h$ must satisfy the condition $\overline\omega^2\approx\overline\omega_\eta A_2$ near the frequency $\overline\omega^2\approx 1+B_h$. That requires that $A_h$ is close to $(1+B_h)/\overline\omega_\eta$ at $\gamma=1$. By recalling the definition of $A_2$ and $B_2$ together with that of $g$, we have
\begin{align}
\frac{\beta \overline\omega_\eta Nq}{4 \omega_r C_0}\left( i_0aN(1-\overline{K}_e)g_x-\frac{\sigma q}{C_0}\right) \approx 1+\frac{\beta Nq^2}{4C_0}.
\end{align}
assuming $u_a$ is small (See numerical examination below).

Since the capacitance $C_0$ and the number $N$ of the motile protein are both approximately proportional to the cell length, the ratio $N/C_0$ is relatively conserved. Thus the right-hand-side of this equation is positive and is approximately constant. This requirement has the following consequences.

First, to satisfy this equality, the current $i_0$ needs to be larger if the external elastic load increases and making $\overline K_e$ approach unity. That is the case for the basal turn of the cochlea, where the stiffness of shorter OHCs cannot match the stiffness of the basilar membrane. Second, this equality at a higher resonance frequency $\omega_r$ can be maintained only by an increased $i_0$. Thus, this equality indicates a limit of the effectiveness of OHCs as an amplifier in this mode of motion. More details of this condition will be discussed later in the numerical section.

Eqs.\ \ref{eq:HFreq} and \ref{eq:HFreqAB} predict that hyperpolarization leads to reduced effectiveness of the OHC because it is expected to reduce $i_o$ and reduces $\gamma$, away from the mid point ($\sim -50$ mV) of conformational transition. The effect of depolarization is less clear because it increases $i_0$ and decreases $\gamma$ at the same time. The outcome depends on the balance of the two.

\subsection*{Low frequency approximation} 
If the frequency $\omega$ is low enough to satisfy $\omega C_0/\sigma\ll 1$, i.e.\ the resistive conductance is larger than the capacitive conductance, an approximate form for low frequency could be formed by replacing $1/(\sigma+i\omega C_0)$ with $(1/\sigma)(1-i\omega C_0/\sigma)$. 

However, a much simpler form is provided by simply nullifying $C_0$  because $\sigma/(\omega_r C_0)$ is quite small as shown later in the numerical section and the frequency range this from applies is quite narrow. 

With this replacement, Eq.\ \ref{eq:complex_eom1} turns into
\begin{align}
[-\overline\omega^2+i\overline\omega(1/\overline\omega_\eta+\gamma A_l)+1+\gamma B_l] p=(1+\alpha_c\gamma u_a)\overline{f}
\label{eq:LFreq}
\end{align}
with a damping factor $A_1$ and a stiffness factor $B_1$, which are respectively defined by
\begin{subequations}
\label{eq:LFreqAB}
\begin{align}
A_l&=\frac{\beta\omega_rNq^2}{4\sigma}, \\
B_l&=-u_a\overline K_e+\frac{\beta  gi_0 q}{4\sigma},\\
\end{align}
\end{subequations}
The coefficient $A_L$, which originates from strain-induced polarization, is positive, indicating that the piezoelectric lateral wall produces powerful damping force.  It also shows that hair bundle conductance increases the stiffness of OHCs. 


\subsection*{Gating compliance} 

Now consider the low frequency asymptote under a constant voltage condition.  If we let $\omega\rightarrow 0$ and impose constant voltage, Eq.\ \ref{eq:complex_eom1} turns into
\begin{align}
(1-\gamma u_a\overline K_e) p=(1+\alpha_c\gamma u_a)\overline{f}.
\label{eq:p_v_f}
\end{align}
In the absence of external load, $\overline K_e=0$ and $\alpha_c=1$.
Recalling the relationship between $x$ and $p$ and that of between $F_\mathrm{ext}$ and $f$, we obtain compliance as
\begin{align}
\frac{x}{F_\mathrm{ext}}=\frac{1+\gamma u_a}{k_o}.
\label{eq:gating_compliance}
\end{align}
The reason for ``gating'' compliance  is because the parameter $u_a$, which is proportional to $\gamma$ (see Eq.\ \ref{eq:ua_def}), maximizes at $P_0=0.5$, where the motile element undergoes the sharpest conformational changes in response to changes in the external force $F_\mathrm{ext}$. However, this effect is  minor because $u_a$ is small as shown in the next section.

\section*{Numerical examination} 
Here the performance of a mid-range frequency OHC in the guinea pig cochlea is examined as an example. The cell length is 40 $\mu$m, and the structural capacitance $C_0$ is 30 pF and the location of the cell is 4 kHz. 

\subsubsection*{Parameter values}

The values used are listed in Table \ref{tab:param_vals}. An OHC near the 4 kHz location in the guinea pig cochlea has on  average the membrane capacitance of 30 pF and the membrane resistance of 100 M$\Omega$ \cite{ma1996}. If we assume that the reversal potential of the basolateral membrane is close to $-80$ mV, the Nernst potential for K$^+$, the steady state current $i_0$ is 0.3 nA for the resting potential of $-50$ mV (steady state current of Eq.\ \ref{eq:Ra}), and 0.4 nA for the more depolarized $-40$ mV \cite{Johnson2011}.

\begin{table}[h]
\caption{Parameter values used for plots. The values chosen are for 40 $\mu$m OHC and is assumed to corresponds to the 4 kHz location in guinea pigs.}
\begin{center}
\begin{tabular}{|c|c|r|c|r|}
\hline\hline
notation & definition & value &  ref. \\
\hline
$q$ & unit motile charge & 0.8 $e$ & \cite{i2010}\\
$a$ & unit displacement & $0.67 \times 10^{-4}$ nm & see text \\
$N$ & number of units & $3\times 10^{7}$ & see text \\
$C_0$ & structural capacitance & 30 pF & \cite{ma1996}\\
$\sigma$ & membrane conductance & 10 nS &\cite{ma1996} \\
$i_0$ & baseline current & 0.3 nA &  see text \\
$k_o$ & OHC stiffness & 17 mN/m & \cite{ia1997} \\
$K_e$ & BM stiffness & 20 mN/m & \cite{Gummer1987},\cite{olson2012} \\
$g_x$ & HB sensitivity & 1/(25 nm) &  \cite{rrc1986} \\
$\eta$ & drag coefficient & $0.8\times 10^{-7}$N/m & see text \\
\hline
\end{tabular}
\end{center}
\caption*{Note: $e$ is the electronic charge of $1.6\times 10^{-19}$ C. BM: the basilar membrane, HB: hair bundle.
}
\label{tab:param_vals}
\end{table}%

The elastic modulus of the guinea pig OHC is  510 nN per unit strain \cite{ia1997}. Because the basal 10 $\mu$m of OHC's are held by the Deiters' cup, the exposed part of a 40 $\mu$m-long OHC is 30 $\mu$m. Thus the estimated stiffness is 17 mN/m (510 nN/30$\mu$m). 

The stiffness of the guinea pig BM is 0.5 N/m at 2 mm from the base, using a probe with a tip diameter of 25 $\mu$m \cite{Gummer1987}. Stiffness of the BM reduces 8 fold from 2 mm from the base to 6 mm from the base, where the characteristic frequency  is 4 kHz \cite{olson2012}. That leads to 60 mN/m. If we can assume somewhat arbitrarily that the stiffness measured corresponds to 3 OHCs, the stiffness of the BM per OHC is 20 mN/m.

The peak excess capacitance due to the motile charge $q$ is $\beta q^2 N/4$. If it is 30 pF, about the same as $C_0$ \cite{skkkt1998}, we obtain $N=3\times 10^7$, assuming $q=-0.8e$. The maximal load-free amplitude of electromotility is $aN$. If we assume it is 5\% of the total length \cite{i2010}, $a=0.67\times10^{-4}$ nm.

The main contribution to the drag can be the shear drag of the subtectorial space \cite{allen1980}.  Then, the drag coefficient $\eta$ is proportional to $S/d$, where $S$ is the gap area per OHC and $d$ is the tallest stereocilia in the hair bundle. If we can assume the area is 10 $\mu$m $\times$ 20 $\mu$m, the gap is the tallest of row 2 $\mu$m \cite{lim1980} of the hair bundle, we obtain $\eta=0.8\times 10^{-7}$ N/m, using the viscosity of water \cite{odi2003a}. For the resonance frequency of 4 kHz, this value leads to $\omega_\eta/\omega_r=10$. If we use this drag coefficient for displacement $x$, assuming this subtectorial shear and hair bundle displacement is the same as BM displacement $x$ \cite{odi2003a}. The sensitivity of hair bundle {resistance}   to $x$ is also based on this assumption.

\subsubsection*{Magnitude of $u_a$}
\begin{SCfigure}[0.8] 
\includegraphics[width=0.65\linewidth]{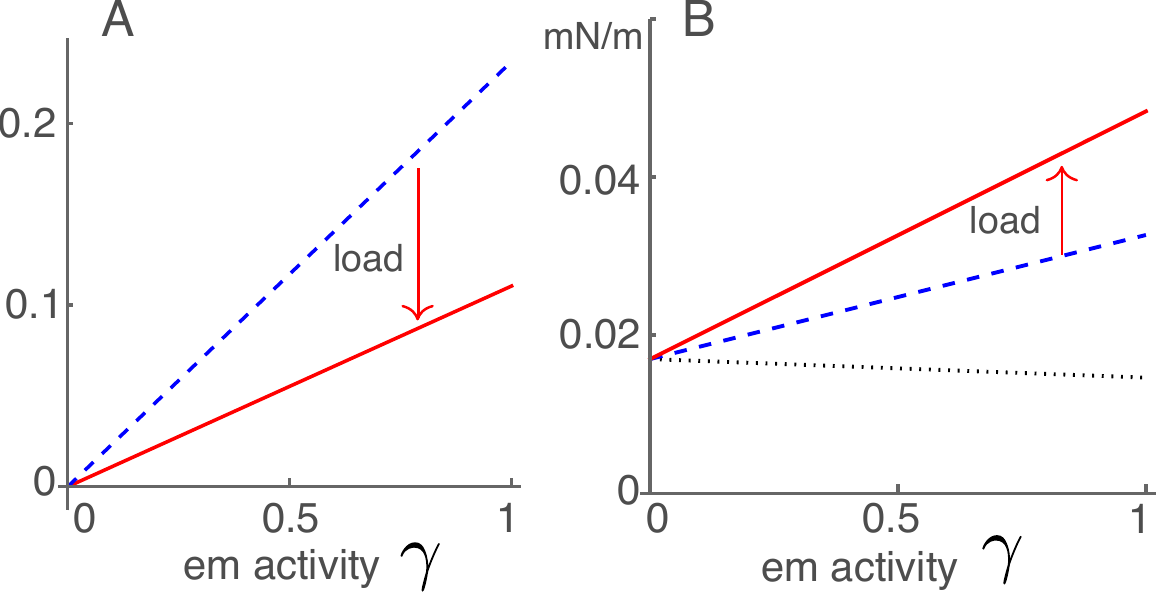}
\caption{\small{Anti-drag coefficient and OHC stiffness. A: Anti-drag term with the elastic load (red solid line) and without elastic load (blue dashed line). Both are proportional to em activity $\gamma$. B: OHC stiffness. With the elastic load (red line), without elastic load (dashed blue line), and isolated and under voltage-clamped condition (dotted line). 
}}
\label{fig:AandStiffness}
\end{SCfigure}

The quantity $u_a(=\beta a^2 Nk_o/4)$ represents the effect of external force on the electromotile element of OHCs. It is also related to the magnitude of gating compliance (Fig.\ \ref{fig:AandStiffness}), and contributes to $B_1$ and $B_2$. It is important in the distinction of the two connectivities A and B (Fig.\ \ref{fig:mech_schem}). The parameter values in Table \ref{tab:param_vals} leads to a value $u_a=0.14$. 

\subsubsection*{Parameters of high frequency approximation}
The equation of motion (Eq.\ \ref{eq:complex_eom1}) can take a simplified form in the region $\omega_r\gg\sigma/C_0$ (high frequency approximation) or $\omega_r\ll\sigma/C_0$ (low frequency approximation) as described earlier. Because the present set of parameter values $\sigma/(\omega_r C_0)=0.013$, the validity of the low frequency approximation is limited to frequencies below 100 Hz.

The values of the parameters in the high frequency approximation (Eq.\ \ref{eq:HFreq}) are
\begin{subequations}
\begin{align} \nonumber
A_h&= \frac{\beta q}{4 \omega_r C_0}\left( i_0 g-\frac{\sigma q}{C_0}\right)\\
&=0.165-0.013=0.152, \\ \nonumber
B_h&=-u_a \overline K_e+\frac{\beta Nq}{4 C_0}\left(q+\frac{\sigma a g i_0}{\omega_r^2\overline\omega^2C_0}\right) \\
&= -0.069 +0.989+0.002(\overline\omega=1)=0.922.
\end{align}
\end{subequations}
Anti-damping term $A_h$ consists of two terms. Even though the first term is about ten times larger than the second, the effect of the second term is still important (See Fig.\ \ref{fig:HF}B1). Because the first term is proportional $(1-\overline K_e)$ (see Eq.\ \ref{eq:g}), $(1-\overline K_e)$ decreases as the ratio $K_e/k_o$ increases with the increase of BM stiffness toward the base. 

The stiffness term $B_h$ has a contribution of gating compliance in the first term. However it is overwhelmed by the second term, which arises from stiffening by strain-induced polarization of the OHC membrane (the last term in Eq.\ \ref{eq:c_eqn}).

\subsubsection*{Parameters of low frequency approximation}
Even though the applicability of this approximation is limited to extremely low frequencies, it is interesting to note which factors contribute to drag and stiffness.
\begin{subequations}
\begin{align} \nonumber
A_l&= \frac{\beta\omega_rNq^2}{4\sigma}=862, \\ \nonumber
B_l&=-u_a\overline K_e+\frac{\beta  gi_0 q}{4\sigma}\\
&=-0.069+12.429 =12.360.
\end{align}
\end{subequations}
Notice that the value of $A_l$ is positive and quite large and that originates from piezoelectricity. Gating compliance is also overwhelmed by the factor, which is coupled with hair bundle {resistance}   in this case.

\subsection*{Amplitude and phase at high frequencies}

\begin{figure}[h!]
\begin{center}
\includegraphics[width=0.7\linewidth]{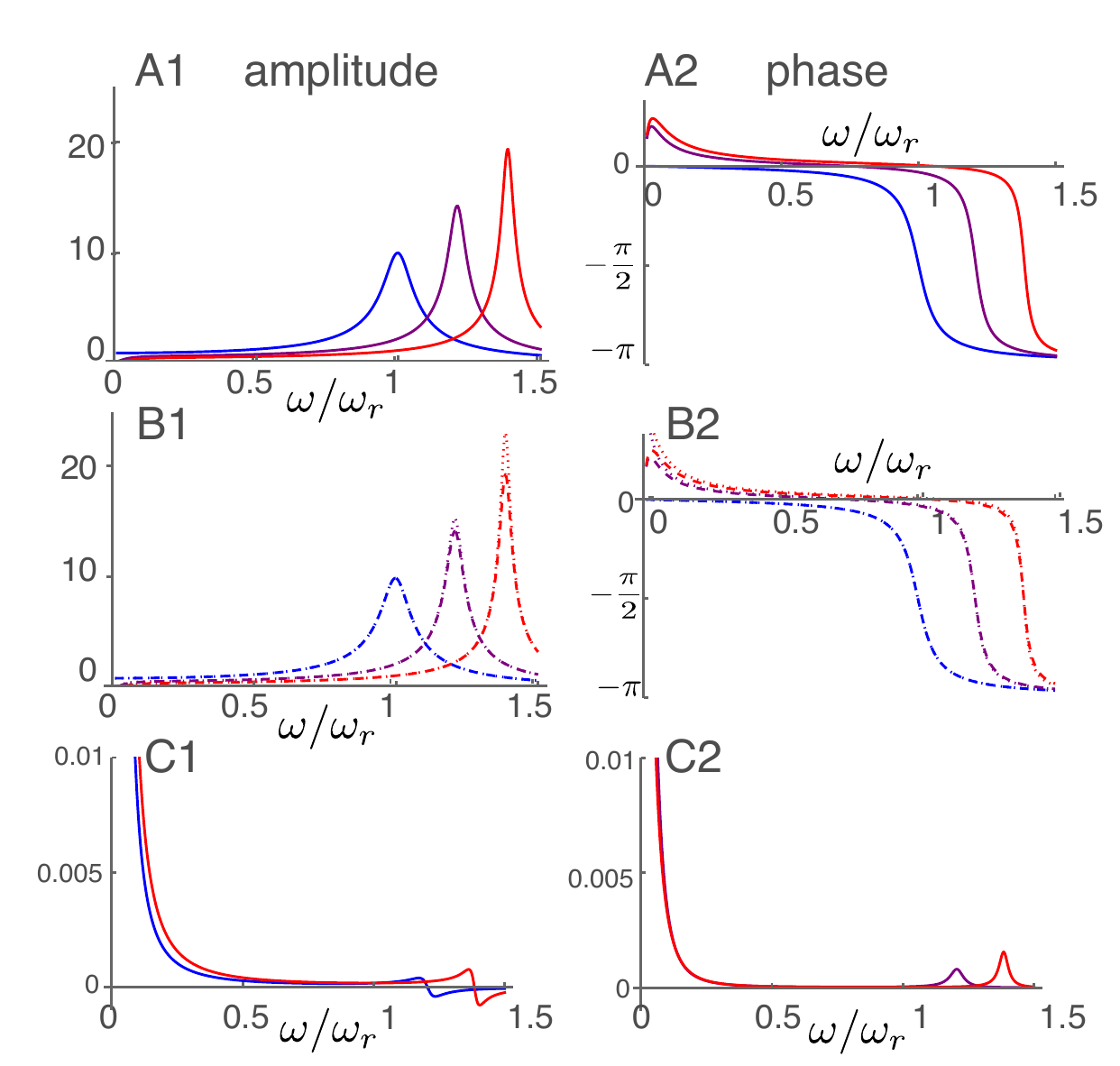}
\caption{\small{Amplitude and phase in the high frequency region plotted against frequency normalized to the resonance frequency. A: Amplitude (A1) and phase (A2) of $p$ calculated from non-expanded form (Eq.\ \ref{eq:complex_eom1}) The unit of amplitude is $\overline f$ and $\alpha_c=1$ is used. B: Amplitude (B1) and phase (B2) of high frequency expansion (Eq.\ \ref{eq:HFreq}). With membrane conductance $\sigma$ (dashed) and without membrane conductance (dotted). Three plots in each corresponds to $\gamma$ values of 0 (blue), 1/2 (purple) and 1 (red), respectively.
C: Difference between non-expanded equation and expanded equation with $\sigma$. Relative difference of amplitude (C1) and difference in phase (C2). No difference for $\gamma=0$.}
}
\label{fig:HF}
\end{center}
\end{figure} 
The frequency dependence of the displacement for the parameter values is plotted in Fig.\ \ref{fig:HF}. The frequency axis is normalized to the resonance frequency. As the activity parameter $\gamma$ of the motile units increases, the peak of amplitude shifts to higher frequencies, reflecting increasing OHC stiffness. The high frequency approximation (right) shows an overall resemblance to the exact form (left) because the middle frequency ($\overline\omega_m$) is 0.1. 

The height of the peak amplitude of Eq.\ \ref{eq:complex_eom1} increases, as expected,  with $\gamma$, the electromotive activity, increases from null to unity (Fig.\ \ref{fig:HF} A1). At the same time, the amplitude peak shifts to higher frequencies. The phase drops by about $\pi$, respectively, near its corresponding amplitude peak. 

The high frequency expansion (Eq.\ \ref{eq:HFreq}) quite well reproduces (dashed lines) both the amplitude (B1) and the phase (B2). If the membrane conductance is nullified the peak hight increases particularly for higher $\gamma$ (dotted lines), showing the effect of the second term in $A_h$ (Eq.\ \ref{eq:HFreqAB}).

The difference between the unexpanded equation and the approximate equation is quite small except for the low frequency region $\omega/\omega_r<0.3$ ,where the approximate form is not valid. The maximal relative error is less than 0.0008 in amplitude. The error in phase is up to 0.001 radian.

\subsection*{Amplitude and phase at low frequencies}
\begin{figure}[h!]
\begin{center}
\includegraphics[width=0.7\linewidth]{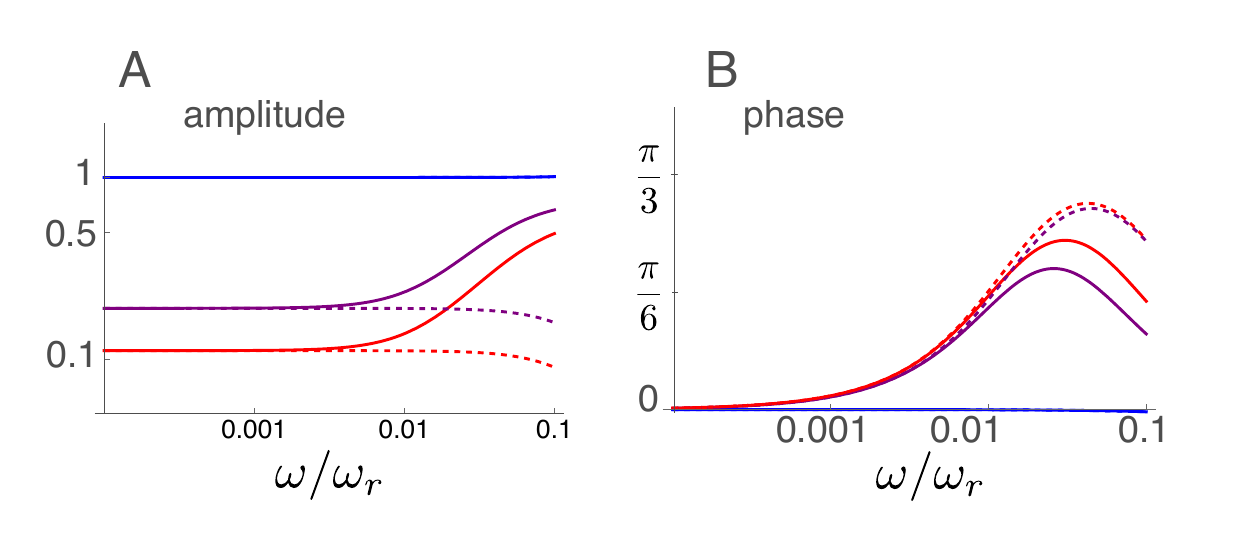}
\caption{\small{Amplitude and phase in the low frequency region. Solid lines: amplitude (A) and phase (B) of the exact form (Eq.\ \ref{eq:complex_eom1}). Dashed lines: amplitude (A) and phase (B) of the appriximated form (Eq.\ \ref{eq:LFreq}). Three lines correspond to $\gamma$ values of 0 (blue), 1/2 (purple), and 1 (red). The unit of amplitude is $\overline f$.
}}
\label{fig:LF}
\end{center}
\end{figure}

The frequency dependence of the displacement at low frequencies is plotted in Fig.\ \ref{fig:LF}. The amplitude is shown in log-log plot and the phase is in log-linear plot.

At these low frequencies, the amplitude decreases with the electromotive activity of OHCs instead of increasing at higher frequencies. The phase increases with frequency from null if the electromotile activity is turned on. Amplitude increases at frequencies higher than $\sim 0.01$, and phase peaks at a frequency of $\sim 0.03$ then decreases in the non-approximated form, but that is outside of the validity of low frequency approximation.

\section*{Discussion} 

OHCs work as an interplay of two mechanosensitive elements, which are coupled by an electric circuit. For this reason, phase relationships provide key insight into their workings. 

\subsection*{Semi-piezoelectric resonance at high frequencies}
\begin{figure}[h!] 
\begin{center}
\includegraphics[width=0.75\linewidth]{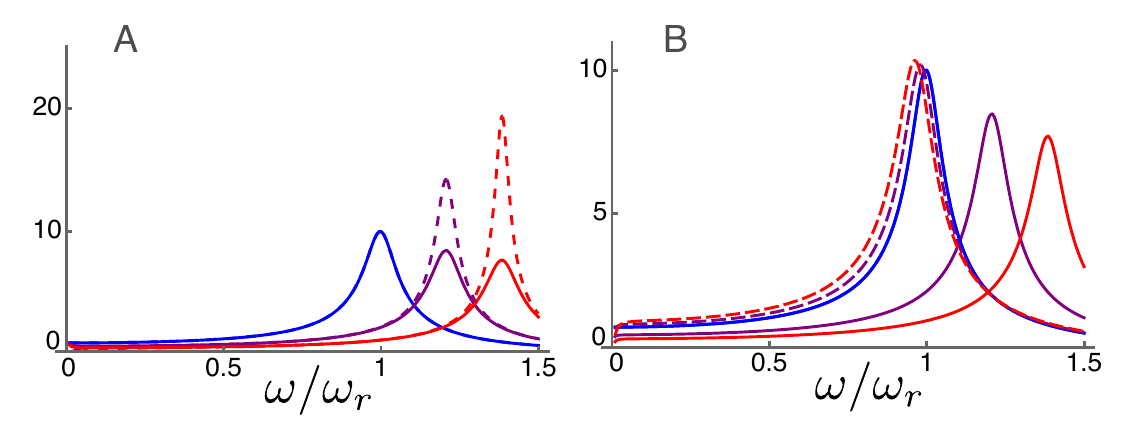}
\caption{\small{Amplitude of pure piezoelectric resonance. A: Amplitude is plotted against frequency with (dashed lines) and without (solid lines) mechanotransducer in the hair bundle. The $\gamma$ values are, respectively, 0 (left), 0.5 (middle), and 1 (right). B: Enlarged plot with (solid lines) and without (dashed lines) strain-induced polarization stiffness (the second term in $B_h$). The $\gamma$ values are, respectively, 0 (right), 0.5 (middle), and 1 (left) for the dashed plots. A small amplitude increase and downward frequency shift with an increase in $\gamma$.
The frequency scale is normalized to the resonance frequency $\omega_r$. 
}}
\label{fig:woHB}
\end{center}
\end{figure}

The role of the mechanosensitivity of the hair bundle is clarified by removing it by putting $\hat r=0$ in Eq.\ \ref{eq:complex_eom1} (or equivalently $g=0$ in Eqs.\ \ref{eq:HFreqAB} and \ref{eq:LFreqAB}). Each peak of Fig.\ \ref{fig:woHB} represents pure piezoelectric resonance.

Under pure piezoelectric resonance the only amplifying factor is the term $\alpha_c\gamma u_a$ on the right-hand-side of Eq.\ \ref{eq:HFreq}A. This term originates from the mechanosensitivity of the lateral wall. Even though it increases with $\gamma$, the peak height decreases (Fig.\ \ref{fig:woHB}A) because the peak frequency goes up away from the pure mechanical resonance frequency, owing to the strain-induced polarization stiffness of the OHC. To test the effect of frequency shift on amplitude, this stiffness term is removed. Then the amplitude slightly increases with increasing $\gamma$ (Fig.\ \ref{fig:woHB}B). Thus, pure piezoelectric resonance does not provide amplification.

This observation demonstrates that the amplitude gain due to OHCs is not due to pure piezoelectric resonance but semi-piezoelectric resonance, in which mechanoelectric sensitivity of the hair bundle plays a crucial role.

\subsection*{Role reversal at low frequencies}
Even though the range of frequencies, where the condition $\sigma/C_0\ll 1$ applies, is narrow,  this condition can still be of interest. At low frequencies where membrane conductance dominates over capacitive conductance, the roles of the two terms that makes major contribution to anti-damping and stiffness increase because the phases of these terms shift. 

Hair bundle {resistance}   is directly associated with stiffness increase and the effect of induced charge contributes to very large drag. For this reason, the amplitude decreases with electromotile activity $\gamma$ in reversal of its effect at high frequencies.

\subsection*{OHC stiffness}
The present model predicts a significant increase of OHC stiffness as the result of hair bundle sensitivity and external elastic load, overwhelming a reduction of stiffness as an analogue to ``gating compliance'' of the hair bundle \cite{Iwasa2000}. Two experimental reports \cite{hd1999,hallworth2007} on the membrane potential dependence of OHC agree with only minor reduction in the stiffness in the physiological range, even though their reports diverge for large depolarizations.

Stiffness increase of OHCs with their displacement appears consistent with the effectiveness in performing their physiological role of cochlear amplifier. Nonetheless,  external elastic load reduces their efficiency of counteracting drag because it reduces parameter $A_h$ as shown in Fig. \ref{fig:AandStiffness}B.

\section*{Conclusions}
The present model shows that existing parameter values in guinea pig at the location of 4 kHz, which are experimentally determined, are consistent with the expected role of OHC as the cochlear amplifier. It predicts that hyperpolarization of the OHCs leads to reduced efficiency. The effect of depolarization is less clear because two factors compete.

OHCs show complex behavior due to the interplay between two mechanosensitive elements. In the narrow low frequency range, where the angular frequency is smaller than $\sigma/C_0$, electromotile activity significantly contributes to drag and stiffness. In the high frequency range, where the angular frequency is greater than $\sigma/C_0$, HB-driven electromotile activity of OHC counteracts drag and piezoelectric sensitivity of the cell body increases stiffness. As the result, the resonance peak increases and peak frequency shifts upward.

The amplifying effect of OHCs decreases with increasing external elastic load and this feature reduces the effectiveness of OHCs as the cochlear amplifier at the basal part of the cochlea, where the BM is much stiffer than in the location of 4 kHz. This limitation is the result of input impedance mismatch, which can be eased by multiple modes of motion.

The resonance that this system exhibits can be called semi-piezoelectric because the exquisite mechanosensitivity of the hair bundle is critical. Pure piezoelectric resonance does not provide amplification because of its small gain combined with the peak shift of the amplitude away from the mechanical resonance frequency.. 

The dominance of hair bundle sensitivity makes the effect of the connectivity, whether parallel or series, of the OHC to external load rather minor. This result facilitates expansion of the present method to more complex systems with multiple degrees of motion.

\nolinenumbers
\section*{Acknowledgments}
This research was supported in part by the Intramural Research Program of the NIH, NIDCD. The author is grateful to Drs.\ Catherine Weisz and Inna Belyantseva for useful comments.

\section*{Declaration of interests}
The author declares no competing interests.


\begin{thebibliography}{10}
\expandafter\ifx\csname url\endcsname\relax
  \def\url#1{\texttt{#1}}\fi
\expandafter\ifx\csname urlprefix\endcsname\relax\def\urlprefix{URL }\fi
\providecommand{\bibinfo}[2]{#2}
\providecommand{\eprint}[2][]{\url{#2}}

\bibitem{ld1984}
\bibinfo{author}{Liberman, M.~C.} \& \bibinfo{author}{Dodds, L.~W.}
\newblock \bibinfo{title}{Single neuron labeling and chronic cochlear
  pathology. {III}. stereocilia damage and alterations of threshold tuning
  curves}.
\newblock \emph{\bibinfo{journal}{Hearing Res.}} \textbf{\bibinfo{volume}{16}},
  \bibinfo{pages}{55--74} (\bibinfo{year}{1984}).

\bibitem{Dallos2008}
\bibinfo{author}{Dallos, P.} \emph{et~al.}
\newblock \bibinfo{title}{Prestin-based outer hair cell motility is necessary
  for mammalian cochlear amplification.}
\newblock \emph{\bibinfo{journal}{Neuron}} \textbf{\bibinfo{volume}{58}},
  \bibinfo{pages}{333--339} (\bibinfo{year}{2008}).

\bibitem{bbbr1985}
\bibinfo{author}{Brownell, W.}, \bibinfo{author}{Bader, C.},
  \bibinfo{author}{Bertrand, D.} \& \bibinfo{author}{Ribaupierre, Y.}
\newblock \bibinfo{title}{Evoked mechanical responses of isolated outer hair
  cells}.
\newblock \emph{\bibinfo{journal}{Science}} \textbf{\bibinfo{volume}{227}},
  \bibinfo{pages}{194--196} (\bibinfo{year}{1985}).

\bibitem{a1987}
\bibinfo{author}{Ashmore, J.~F.}
\newblock \bibinfo{title}{A fast motile response in guinea-pig outer hair
  cells: the molecular basis of the cochlear amplifier}.
\newblock \emph{\bibinfo{journal}{J. Physiol.}} \textbf{\bibinfo{volume}{388}},
  \bibinfo{pages}{323--347} (\bibinfo{year}{1987}).

\bibitem{hh1987}
\bibinfo{author}{Howard, J.} \& \bibinfo{author}{Hudspeth, A.~J.}
\newblock \bibinfo{title}{Mechanical relaxation of the hair bundle mediates
  adaptation in mechanoelectrical transduction by the bullfrog's saccular hair
  cell}.
\newblock \emph{\bibinfo{journal}{Proc. Natl. Acad. Sci. USA}}
  \textbf{\bibinfo{volume}{84}}, \bibinfo{pages}{3064--3068}
  (\bibinfo{year}{1987}).

\bibitem{hh1988}
\bibinfo{author}{Howard, J.} \& \bibinfo{author}{Hudspeth, A.~J.}
\newblock \bibinfo{title}{Compliance of the hair bundle associated with gating
  of mechanoelectrical transduction channels in the bullfrog's saccular hair
  cell}.
\newblock \emph{\bibinfo{journal}{Neuron}} \textbf{\bibinfo{volume}{1}},
  \bibinfo{pages}{189--199} (\bibinfo{year}{1988}).

\bibitem{zshlmd2000}
\bibinfo{author}{Zheng, J.} \emph{et~al.}
\newblock \bibinfo{title}{Prestin is the motor protein of cochlear outer hair
  cells}.
\newblock \emph{\bibinfo{journal}{Nature}} \textbf{\bibinfo{volume}{405}},
  \bibinfo{pages}{149--155} (\bibinfo{year}{2000}).

\bibitem{ric-fet2003}
\bibinfo{author}{Ricci, A.~J.}, \bibinfo{author}{Crawford, A.~C.} \&
  \bibinfo{author}{Fettiplace, R.}
\newblock \bibinfo{title}{Tonotopic variation in the conductance of the hair
  cell mechanotransducer channel}.
\newblock \emph{\bibinfo{journal}{Neuron}} \textbf{\bibinfo{volume}{40}},
  \bibinfo{pages}{983--990} (\bibinfo{year}{2003}).

\bibitem{BeurgRicci2009}
\bibinfo{author}{Beurg, M.}, \bibinfo{author}{Fettiplace, R.},
  \bibinfo{author}{Nam, J.} \& \bibinfo{author}{Ricci, A.}
\newblock \bibinfo{title}{Localization of inner hair cell mechanotransducer
  channels using high-speed calcium imaging}.
\newblock \emph{\bibinfo{journal}{Nat. Neurosci.}}
  \textbf{\bibinfo{volume}{12}}, \bibinfo{pages}{553--558}
  (\bibinfo{year}{2009}).

\bibitem{a1990}
\bibinfo{author}{Ashmore, J.~F.}
\newblock \bibinfo{title}{Forward and reverse transduction in guinea-pig outer
  hair cells: the cellular basis of the cochlear amplifier}.
\newblock \emph{\bibinfo{journal}{Neurosci. Res. Suppl.}}
  \textbf{\bibinfo{volume}{12}}, \bibinfo{pages}{S39--S50}
  (\bibinfo{year}{1990}).

\bibitem{i1993}
\bibinfo{author}{Iwasa, K.~H.}
\newblock \bibinfo{title}{Effect of stress on the membrane capacitance of the
  auditory outer hair cell}.
\newblock \emph{\bibinfo{journal}{Biophys. J.}} \textbf{\bibinfo{volume}{65}},
  \bibinfo{pages}{492--498} (\bibinfo{year}{1993}).

\bibitem{ia1997}
\bibinfo{author}{Iwasa, K.~H.} \& \bibinfo{author}{Adachi, M.}
\newblock \bibinfo{title}{Force generation in the outer hair cell of the
  cochlea}.
\newblock \emph{\bibinfo{journal}{Biophys. J.}} \textbf{\bibinfo{volume}{73}},
  \bibinfo{pages}{546--555} (\bibinfo{year}{1997}).

\bibitem{Santos-Sacchi1998a}
\bibinfo{author}{Santos-Sacchi, J.}, \bibinfo{author}{Kakehata, S.} \&
  \bibinfo{author}{Takahashi, S.}
\newblock \bibinfo{title}{Effects of membrane potential on the voltage
  dependence of motility-related charge in outer hair cells of the guinea-pig.}
\newblock \emph{\bibinfo{journal}{J Physiol}} \textbf{\bibinfo{volume}{510 ( Pt
  1)}}, \bibinfo{pages}{225--235} (\bibinfo{year}{1998}).

\bibitem{ai1999}
\bibinfo{author}{Adachi, M.} \& \bibinfo{author}{Iwasa, K.~H.}
\newblock \bibinfo{title}{Electrically driven motor in the outer hair cell:
  Effect of a mechanical constraint}.
\newblock \emph{\bibinfo{journal}{Proc. Natl. Acad. Sci. USA}}
  \textbf{\bibinfo{volume}{96}}, \bibinfo{pages}{7244--7249}
  (\bibinfo{year}{1999}).

\bibitem{SantosNava2023}
\bibinfo{author}{Santos-Sacchi, J.}, \bibinfo{author}{Bai, J.-P.} \&
  \bibinfo{author}{Navaratnam, D.}
\newblock \bibinfo{title}{Megahertz sampling of prestin ({SLC}26a5)
  voltage-sensor charge movements in outer hair cell membranes reveals
  ultrasonic activity that may support electromotility and cochlear
  amplification}.
\newblock \emph{\bibinfo{journal}{J. Neurosci.}} \textbf{\bibinfo{volume}{14}},
  \bibinfo{pages}{2460--2468} (\bibinfo{year}{2023}).

\bibitem{i1994}
\bibinfo{author}{Iwasa, K.~H.}
\newblock \bibinfo{title}{A membrane model for the fast motility of the outer
  hair cell}.
\newblock \emph{\bibinfo{journal}{J. Acoust. Soc. Am.}}
  \textbf{\bibinfo{volume}{96}}, \bibinfo{pages}{2216--2224}
  (\bibinfo{year}{1994}).

\bibitem{i1990}
\bibinfo{author}{Ikeda, T.}
\newblock \emph{\bibinfo{title}{Fundamentals of Piezoelectricity}}
  (\bibinfo{publisher}{Oxford University Press}, \bibinfo{address}{Oxford, UK},
  \bibinfo{year}{1990}).

\bibitem{Iwasa2021}
\bibinfo{author}{Iwasa, K.~H.}
\newblock \bibinfo{title}{Kinetic membrane model of outer hair cells}.
\newblock \emph{\bibinfo{journal}{Biophys. J.}} \textbf{\bibinfo{volume}{120}},
  \bibinfo{pages}{122--132} (\bibinfo{year}{2021}).

\bibitem{ma1996}
\bibinfo{author}{Mammano, F.} \& \bibinfo{author}{Ashmore, J.~F.}
\newblock \bibinfo{title}{Differential expression of outer hair cell potassium
  currents in the isolated cochlea of the guinea-pig}.
\newblock \emph{\bibinfo{journal}{J. Physiol.}} \textbf{\bibinfo{volume}{496}},
  \bibinfo{pages}{639--646} (\bibinfo{year}{1996}).

\bibitem{Johnson2011}
\bibinfo{author}{Johnson, S.~L.}, \bibinfo{author}{Beurg, M.},
  \bibinfo{author}{Marcotti, W.} \& \bibinfo{author}{Fettiplace, R.}
\newblock \bibinfo{title}{Prestin-driven cochlear amplification is not limited
  by the outer hair cell membrane time constant.}
\newblock \emph{\bibinfo{journal}{Neuron}} \textbf{\bibinfo{volume}{70}},
  \bibinfo{pages}{1143--1154} (\bibinfo{year}{2011}).

\bibitem{i2010}
\bibinfo{author}{Iwasa, K.~H.}
\newblock \bibinfo{title}{Chapter 6. {E}lectromotility of outer hair cells}.
\newblock In \bibinfo{editor}{Fuchs, P.~A.} (ed.) \emph{\bibinfo{booktitle}{The
  {O}xford {H}andbook of {A}uditory {S}cience volume 1: {T}he {E}ar}},
  \bibinfo{pages}{179--212} (\bibinfo{publisher}{Oxford University Press},
  \bibinfo{address}{Oxford, UK}, \bibinfo{year}{2010}).

\bibitem{Gummer1987}
\bibinfo{author}{Gummer, A.~W.}, \bibinfo{author}{Smolders, J.~W.} \&
  \bibinfo{author}{Klinke, R.}
\newblock \bibinfo{title}{Basilar membrane motion in the pigeon measured with
  the {M}\"{o}ssbauer technique.}
\newblock \emph{\bibinfo{journal}{Hear Res}} \textbf{\bibinfo{volume}{29}},
  \bibinfo{pages}{63--92} (\bibinfo{year}{1987}).

\bibitem{olson2012}
\bibinfo{author}{Olson, E.~S.}, \bibinfo{author}{Duifhuis, H.} \&
  \bibinfo{author}{Steele, C.~R.}
\newblock \bibinfo{title}{Von b\'{e}k\'{e}sy and cochlear mechanics}.
\newblock \emph{\bibinfo{journal}{Hearing Res.}}
  \textbf{\bibinfo{volume}{213}}, \bibinfo{pages}{31--43}
  (\bibinfo{year}{2012}).

\bibitem{rrc1986}
\bibinfo{author}{Russell, I.~J.}, \bibinfo{author}{Richardson, G.~P.} \&
  \bibinfo{author}{Cody, A.~R.}
\newblock \bibinfo{title}{Mechanosensitivity of mammalian auditory hair cells
  \textit{in vitro}}.
\newblock \emph{\bibinfo{journal}{Nature}} \textbf{\bibinfo{volume}{321}},
  \bibinfo{pages}{517--519} (\bibinfo{year}{1986}).

\bibitem{skkkt1998}
\bibinfo{author}{Santos-Sacchi, J.}, \bibinfo{author}{Kakehata, S.},
  \bibinfo{author}{Kikuchi, T.}, \bibinfo{author}{Katori, Y.} \&
  \bibinfo{author}{Takasaka, T.}
\newblock \bibinfo{title}{Density of motility-related charge in the outer hair
  cell of the guinea pig is inversely related to best frequency}.
\newblock \emph{\bibinfo{journal}{Neurosci Lett.}}
  \textbf{\bibinfo{volume}{256}}, \bibinfo{pages}{155--158}
  (\bibinfo{year}{1998}).

\bibitem{allen1980}
\bibinfo{author}{Allen, J.}
\newblock \bibinfo{title}{Cochlear micromechanics---a physical model of
  transduction}.
\newblock \emph{\bibinfo{journal}{J. Acoust. Soc. Am.}}
  \textbf{\bibinfo{volume}{68}}, \bibinfo{pages}{1660--1670}
  (\bibinfo{year}{1980}).

\bibitem{lim1980}
\bibinfo{author}{Lim, D.~J.}
\newblock \bibinfo{title}{Cochlear anatomy related to cochlear micromechanics.
  a review}.
\newblock \emph{\bibinfo{journal}{J. Acoust. Soc. Am.}}
  \textbf{\bibinfo{volume}{67}}, \bibinfo{pages}{1686--1695}
  (\bibinfo{year}{1980}).

\bibitem{odi2003a}
\bibinfo{author}{Ospeck, M.}, \bibinfo{author}{Dong, X.-X.} \&
  \bibinfo{author}{Iwasa, K.~H.}
\newblock \bibinfo{title}{Limiting frequency of the cochlear amplifier based on
  electromotility of outer hair cells}.
\newblock \emph{\bibinfo{journal}{Biophys. J.}} \textbf{\bibinfo{volume}{84}},
  \bibinfo{pages}{739--749} (\bibinfo{year}{2003}).

\bibitem{Iwasa2000}
\bibinfo{author}{Iwasa, K.~H.}
\newblock \bibinfo{title}{Effect of membrane motor on the axial stiffness of
  the cochlear outer hair cell.}
\newblock \emph{\bibinfo{journal}{J Acoust Soc Am}}
  \textbf{\bibinfo{volume}{107}}, \bibinfo{pages}{2764--2766}
  (\bibinfo{year}{2000}).

\bibitem{hd1999}
\bibinfo{author}{He, D. Z.~Z.} \& \bibinfo{author}{Dallos, P.}
\newblock \bibinfo{title}{Somatic stiffness of cochlear outer hair cells is
  voltage-dependent}.
\newblock \emph{\bibinfo{journal}{Proc. Natl. Acad. Sci. USA}}
  \textbf{\bibinfo{volume}{96}}, \bibinfo{pages}{8223--8228}
  (\bibinfo{year}{1999}).

\bibitem{hallworth2007}
\bibinfo{author}{Hallworth, R.}
\newblock \bibinfo{title}{Absence of voltage-dependent compliance in
  high-frequency cochlear outer hair cells.}
\newblock \emph{\bibinfo{journal}{J Assoc Res Otolaryngol}}
  \textbf{\bibinfo{volume}{8}}, \bibinfo{pages}{464--473}
  (\bibinfo{year}{2007}).

\end{thebibliography}

\end{document}